\documentclass[12pt,draftclsnofoot, onecolumn]{IEEEtran}

\usepackage[cmex10]{amsmath}
\usepackage{amsfonts,amssymb,cite,graphicx,delarray}
\usepackage{multirow}
\IEEEoverridecommandlockouts

\newtheorem{property}{Property}
\newtheorem{theorem}{Theorem}

\newtheorem{remark}{Remark}
\newtheorem{lemma}{Lemma}
\newtheorem{proposition}{Proposition}
\begin{document}


\title{Throughput and Energy Efficiency Analysis of Small Cell Networks with Multi-antenna Base Stations}
\date{ }
\author{	
	\IEEEauthorblockN{Chang Li, Jun Zhang, and Khaled B. Letaief, {\it Fellow, IEEE}}
\thanks{The authors are with Dept. of ECE, The Hong Kong University of Science and Technology. Email: \{changli, eejzhang, eekhaled\}@ust.hk.} }

\begin{titlepage}
\maketitle
\thispagestyle{empty}
\vspace{-1.5cm}
\begin{abstract}
    Small cell networks have recently been proposed as an important evolution path for the next-generation cellular networks. However, with more and more irregularly deployed base stations (BSs), it is becoming increasingly difficult to quantify the achievable network throughput or energy efficiency. In this paper, we develop an analytical framework for downlink performance evaluation of small cell networks, based on a random spatial network model, where BSs and users are modeled as two independent spatial Poisson point processes. A new simple expression of the outage probability is derived, which is analytically tractable and is especially useful with multi-antenna transmissions. This new result is then applied to evaluate the network throughput and energy efficiency. It is analytically shown that deploying more BSs or more BS antennas can always increase the network throughput, but the performance gain critically depends on the BS-user density ratio and the number of BS antennas. On the other hand, increasing the BS density or the number of transmit antennas will first increase and then decrease the energy efficiency if different components of BS power consumption satisfy certain conditions, and the optimal BS density and the optimal number of BS antennas can be found. Otherwise, the energy efficiency will always decrease. Simulation results shall demonstrate that our conclusions based on the random network model are general and also hold in a regular grid-based model.
\end{abstract}

\vspace{-0.5cm}
\begin{keywords}
\vspace{-0.5cm} Poisson point process, stochastic geometry, cellular networks, outage probability, network throughput, energy efficiency.
\end{keywords}

\end{titlepage}

\baselineskip=24pt
\newpage

\section{Introduction}

In recent years, significant technological advances have occurred and have been used to improve capacity and the performance of wireless networks.  Such innovations are significant and have already brought spectral efficiency of point-to-point communication links close to theoretical limits. Unfortunately, if we wish to meet the stringent demands of next-generation wireless systems and beyond 4G networks, new and revolutionary ways have to be developed to address the projected significant increase in mobile data traffic.   Heterogeneous networks, where small base stations (BSs) are overlaid within the macro network based on traffic/coverage demand, represent a new paradigm for significantly expanding network capacity as well as an attractive cost-effective solution for providing a uniform user experience \cite{Hoydis11}.
Different types of BSs will be deployed, forming macro cells, micro cells, pico-cells, and femto-cells \cite{Andrews12}. Meanwhile, lots of BSs will be deployed by the end users, which makes the network more irregular. All these factors bring difficulties to evaluating the achievable throughput gain by densifying the network. In the meantime, green communications is drawing more and more attention on a global scale, and higher energy efficiency is among the main design objectives of the next-generation cellular networks. In \cite{Hasan11}, it was pointed out that BSs consume more than $60\%$ of the total energy in cellular networks. As more and more BSs are deployed, the effect on the energy efficiency should be carefully investigated. In this paper, we endeavor to develop an analytical framework for evaluating both the network throughput and energy efficiency in small cell networks, as well as provide design guidelines for practical deployment.

\subsection{Related Works and Motivation}

Previous works that investigate the throughput, energy efficiency and their tradeoff have mainly focused on the point-to-point communication link or the single-cell case \cite{Deng13,Li12,Isheden12,Xiong11}, while the interference from other cells are neglected. Meanwhile, the throughput analysis of conventional cellular networks has received lots of attention, and different models have been proposed, such as the Wyner model \cite{Wyner94} or the grid model \cite{Ramanath09,Kurniawan12}. While the Wyner model is commonly used due to its tractability, it may lose the essential characteristics of real and practical networks \cite{Xu11}. On the other hand, the regular grid model becomes intractable as the network size grows, and it cannot handle the irregular network structure in small cell networks. In general, it is quite challenging to accurately evaluate the performance of cellular networks, due to the complexity of the network topology, and effects of multi-path propagation. A more common way to evaluate cellular networks is by simulation. For example, in \cite{Kurniawan12}, different cellular network architectures were compared through simulation. While evaluating the network performance through simulation can provide insights on some specific settings, the results may not be extended to other scenarios and the computational complexity is rather high.

Recently, Andrews \emph{et al.} proposed a random spatial model where BSs are modeled as a spatial Poisson point process (PPP) \cite{Andrews11}. Such kind of random network model has been used extensively in wireless ad-hoc networks \cite{Weber05,Baccelli06,Hunter08,Wu12,Louie11}, and it is well suited for small cell networks, where BS positions are becoming irregular. Moreover, with the help of stochastic geometry and the point process theory \cite{Stoyan87,Baccelli09,Haenggi09_SG}, this model has been shown to be tractable and accurate, and can be used to analyze the outage probability and throughput in cellular networks. This random spatial network model has also been used to analyze other networks such as heterogeneous cellular networks \cite{Chandrasekhar09,Cheung12,Cao12,Dhillon12}, distributed antenna systems \cite{Zhang08}, and cognitive radio networks \cite{Lee11,Lee12_CR}.

So far, most studies that adopt the random network model to analyze cellular networks only focus on the spatial distribution of BSs, while the distribution of mobile users is largely ignored. Specifically, BSs are modeled as a PPP, and each BS always has a mobile user to serve, so the user density and the BS-user association are irrelevant. Such an assumption holds only when the user density is much larger than the BS density, which is not the case in small cell networks where the user density is comparable to the BS density. In this paper, we will explicitly consider the user density and the BS-user association. Moreover, most previous works only consider single-antenna BSs. As shown in previous works on wireless ad-hoc networks \cite{Hunter08,Wu12,Louie11}, random network models with multi-antenna transmission are much more challenging than single-antenna systems. In cellular networks, stochastic orders were introduced in \cite{Dhillon13} to provide qualitative comparison between different multi-antenna techniques, but such method cannot be used for quantitative analysis. In our work, we will consider multi-antenna transmission in small cell networks and investigate the effect of multiple BS antennas on the system performance.

\subsection{Contributions}

In this paper, by applying the spatial random network model, we develop a new set of analytical results to evaluate both the network throughput and energy efficiency of small cell networks. We first derive a new simple expression of the outage probability for a typical user, which is expressed in a much simpler form than previous developed results and is more tractable especially with multi-antenna transmissions. It is then used to derive several key properties of the outage probability, demonstrating the effects of the BS density and the number of BS antennas.

Based on the new expression of the outage probability, we evaluate both the network throughput and energy efficiency of small cell networks. For network throughput, a key finding is that while it always increases as the BS density ($\lambda_b$) increases, the BS-user density ratio determines the scaling law. Specifically, for a given user density $\lambda_u$, when $\lambda_b\ll\lambda_u$, as considered in \cite{Andrews11}, the network throughput grows linearly with $\lambda_b$. But if $\lambda_b\sim\lambda_u$, the network throughput appears to grow logarithmically with $\lambda_b$. Therefore, as most previous works assumed that all BSs are active, i.e., $\lambda_b\ll\lambda_u$, these results cannot be applied in small cell networks, where $\lambda_b\sim\lambda_u$ and the BS activity probability needs to be taken into consideration. It is also shown that deploying more BS antennas will increase the network throughput but the gain diminishes when the number of antennas is further increased.

We shall show that the effect of the BS density or the number of BS antennas on the energy efficiency has two patterns: 1) Increasing the BS density or the number of BS antennas can first increase the energy efficiency to a maximal value and then decrease; and 2) Deploying more BSs or more antennas will always decrease energy efficiency. We find that the thresholds for these two patterns depend critically on the different parts of the BS power consumption model. Moreover, for the first pattern, we derive the optimal BS density and the optimal number of BS antennas that maximize energy efficiency.

\subsection{Paper Organization}

The rest of the paper is organized as follows. Section \ref{Sec:SystemModel} presents the system model and the performance metrics we consider in this paper. Section \ref{Sec:Outage} derives a new simple expression of the outage probability and shows its key properties, while in Section \ref{Sec:Throughput}, we evaluate the network throughput and energy efficiency. The simulation results are shown in Section \ref{Sec:NumResults} and Section \ref{Sec:Conclusions} concludes the paper. The key notations and symbols used in the paper
are listed in Table \ref{tab:Notation}.

\begin{table}
    \centering\small
    \caption{\label{tab:Notation}Key notations and symbols used in the paper}
\begin{tabular}{|c|l|}
    \hline
    Symbol & Definition/Explanation \\ \hline
    $\lambda_b$ & BS density \\
    $\lambda_u$ & User density \\
    $\rho$ & BS-user density ratio, i.e., $\frac{\lambda_b}{\lambda_u}$ \\
    $M$ & Number of transmit antennas per BS \\
    $\alpha$ & Pathloss exponent \\
    $\hat{\gamma}$ & SINR threshold \\
    $p_{\rm a}$ & BS activity probability \\
    $p_{\rm out}$ & Outage probability \\
    $p_{\rm s}$ & Successful transmission probability \\
    $R_0$ & Constant transmission rate ($\triangleq\log_2\left(1+\hat{\gamma}\right)$) \\
    $R_{\rm a}$ & Network throughput \\
    $R_{\rm u}$ & Average throughput per user \\
    $\eta_{\rm EE}$ & Network energy efficiency \\
    \hline
\end{tabular}
\end{table}

\section{System model and Network Performance Metrics} \label{Sec:SystemModel}

In this section, we will first describe the random spatial model for modeling small cell networks, and then present the main performance metrics used in the paper.

\subsection{The Network Model}

\begin{figure}
    \begin{center}
    \scalebox{0.45}{\includegraphics{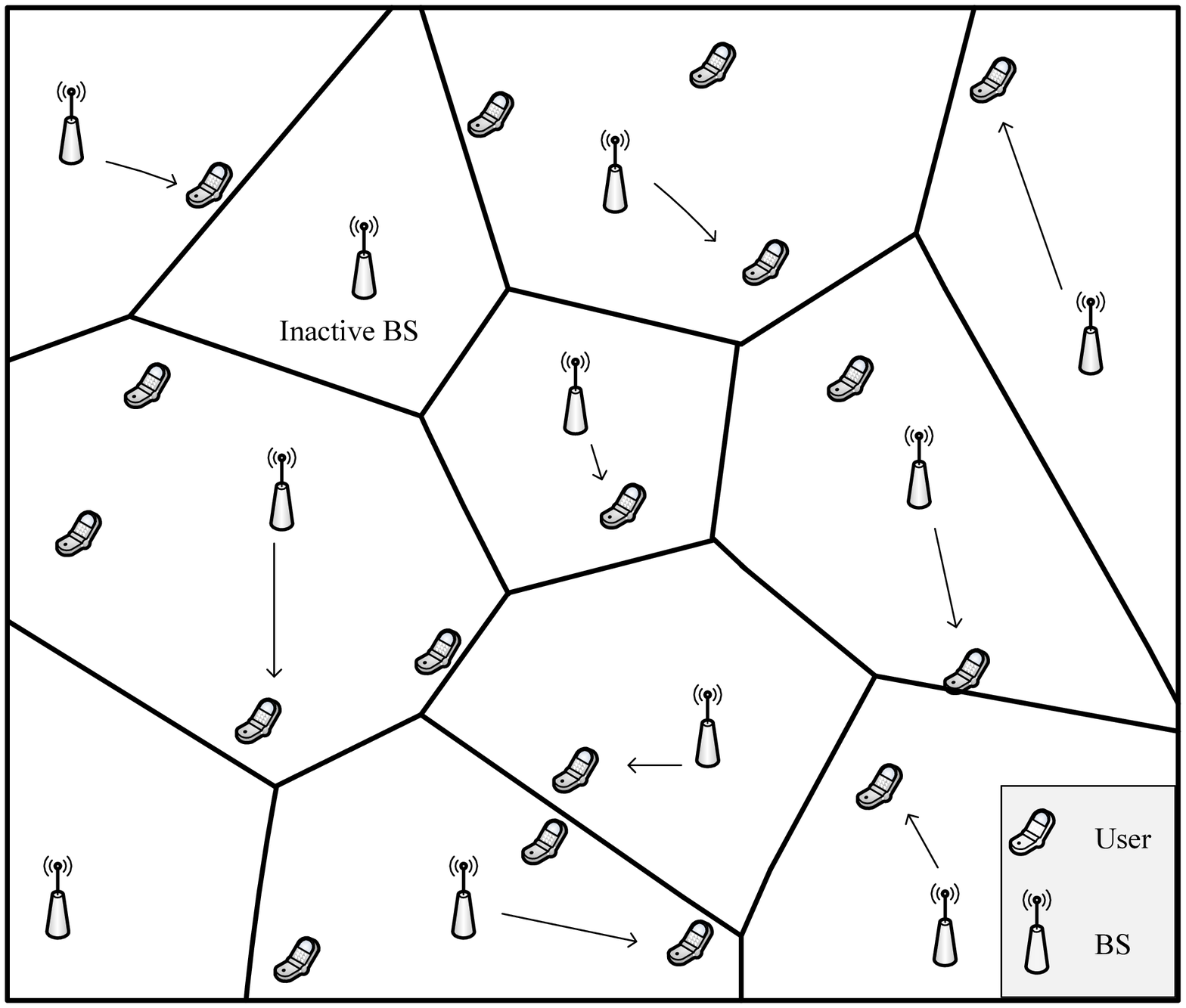}}
    \end{center}
    \caption{A sample network where BSs and users are modeled as two independent PPPs. Each user connects to the closest BS.}
    \label{fig:SystemModel}
\end{figure}

We consider a cellular network, as shown in Fig. \ref{fig:SystemModel}, where BSs and users are distributed according to two independent homogeneous PPPs in $\mathbb{R}^{2}$, denoted as $\Phi_b$ and $\Phi_u$, respectively. Denote the BS density as $\lambda_b$ and the user density as $\lambda_u$. This system can be regarded as a dense deployment of a particular type of BSs or as one tier in a heterogeneous network with orthogonal spectrum allocation among different tiers.

We consider the downlink transmission and assume that each user is served by the nearest BS, which comprises a Voronoi tesselation relative to $\Phi_b$, so the shape of each cell is irregular. This kind of network model is suitable for small cell networks, where BSs are deployed irregularly \cite{Dhillon12,Zhang08}. Due to the independent locations of BSs and users, there may be some BSs that do not have any user to serve. These BSs are called \emph{inactive} BSs and will not transmit any signals, while BSs who have users to serve are called as \emph{active} BSs. The probability that a typical BS is active is denoted as $p_{\rm a}$. Equivalently, $p_{\rm a}$ can be regarded as the ratio of the number of active BSs to the total number of BSs. It has been shown that $p_{\rm a}$, as a function of the BS-user density ratio $\rho\triangleq\frac{\lambda_b}{\lambda_u}$, is given by \cite{Lee12}
\begin{equation} \label{eq:pa}
    p_{\rm a}=1-\left(1+\frac{1}{\mu\rho}\right)^{-\mu},
\end{equation}
where $\mu=3.5$ \cite{Lee12,Ferenc07} is a constant related to the cell size distribution obtained through data fitting\footnote{Note that the value of $\mu$ can be different due to the data fitting, e.g., $\mu=4$ was used in \cite{Kiang66}.}. An active BS may have more than one user in its cell, and the BS will randomly choose one user to serve at each time slot, i.e., intra-cell time division multiple access (TDMA) is adopted in this paper. Note that the derivation can be easily extended to other orthogonal multiple access methods, such as FDMA \cite{Cao12} or SDMA \cite{Kountouris12}.

We assume that each BS is equipped with $M$ antennas
, while each user has a single antenna. Universal frequency reuse is assumed, and thus each user not only receives information from its home BS, but also suffers interference from all the other active BSs. Interference suppression through BS cooperation is not considered as we assume that the backhaul links between different BSs are of very limited capacity and real-time inter-BS information exchange required for cooperation cannot be supported. Moreover, due to the large scale of the small cell network, it is difficult to obtain the global channel state information (CSI) at each BS, so we assume that each active BS only has CSI of the channel to its own user.  While different space-time processing techniques can be applied at each multi-antenna BS \cite{Louie11,Hunter08,Dhillon13}, we will focus on maximal ratio transmission (MRT) beamforming, partly due to its simplicity and partly due to the fact that the optimal usage of multiple antennas in this scenario is unknown.

The received signal for a typical user, denoted as the $0$th user, is given by
\begin{equation} \label{eq:Received_Signal}
	y_{0}=r_{0}^{-\frac{\alpha}{2}}\mathbf{h}_{00}^{\dagger}\mathbf{w}_{0}\sqrt{P_{\rm t}}s_{0} +\sum_{i\neq0}R_{i0}^{-\frac{\alpha}{2}}\mathbf{h}_{i0}^{\dagger}\mathbf{w}_{i}\sqrt{P_{\rm t}}s_{i} +n_{0},
\end{equation}
where $\mathbf{h}_{i0}\sim \mathcal{CN}\left(0,\mathbf{I}\right)$ is an $M\times1$ vector denoting the small scale fading between the $i$th active BS and the $0$th user, $r_0$ is the distance between the $0$th BS to the $0$th user, while $R_{i0}$ is the distance from the $i$th BS to the $0$th user. The pathloss exponent is $\alpha$, the precoding vector is $\mathbf{w}_{i}=\frac{\mathbf{h}_{ii}}{\left\Vert \mathbf{h}_{ii}\right\Vert }$, $P_{\rm t}$ is the transmit power, and $n_0$ denotes the additive white Gaussian noise (AWGN) at the receiver. From \eqref{eq:Received_Signal}, the receive signal-to-interference plus noise ratio (SINR) is given by
\begin{equation} \label{eq:SINR_definition}
	{\rm SINR}=\frac{P_{\rm t}g_{00}r_{0}^{-\alpha}} {\sum_{i\in\tilde{\Phi}_b\backslash0}P_{\rm t}g_{i0}R_{i0}^{-\alpha}+\sigma_{n}^{2}},
\end{equation}
where $\tilde{\Phi}_b$ represents the set of active BSs, and $g_{i0}$ is the equivalent channel gain from the $i$th BS to the $0$th user, i.e., $g_{00}=\left\Vert \mathbf{h}_{00}\right\Vert ^{2}\sim{\rm Gamma}\left(1,M\right)$ and $g_{i0}=\frac{\left\Vert \mathbf{h}_{i0}^{\dagger}\mathbf{h}_{ii}\right\Vert ^{2}}{\left\Vert \mathbf{h}_{ii}\right\Vert ^{2}} \sim{\rm Exp}\left(1\right)$ for $i\neq0$ \cite{Hunter08}. One major difficulty in analyzing \eqref{eq:SINR_definition} is the complicated distribution of $\tilde{\Phi}_b$, which is not a simple homogeneous PPP as in \cite{Andrews11} due to the coupling of the numbers of users in each cell \cite{Lee12}. To simplify the following analysis, we make the same approximation as in \cite{Lee12}, i.e., $\tilde{\Phi}_b$ is assumed to be a homogeneous PPP with density $\lambda_bp_a$. Such approximation has been shown to be accurate in \cite{Lee12}, and we will test it later through simulations. Note that all the following analytical results are exact for a homogeneous $\tilde{\Phi}_b$.

\subsection{Network Performance Metrics}

In this paper, we will focus on two performance metrics: Network throughput and energy efficiency. Assuming fixed-rate transmission, both metrics are determined by the outage probability. Outage happens if the receive SINR falls below a given threshold $\hat{\gamma}$, and the associate outage probability is $p_{\rm out}=\Pr\left( {\rm SINR} \leq \hat{\gamma} \right)$, with SINR given in \eqref{eq:SINR_definition}.

The network throughput, denoted as $R_{\rm a}$, is defined as the average number of successfully transmitted bits per sec$\cdot$Hz$\cdot$unit-area, and is given by \cite{Quek11,Cao12,Wu12}
\begin{equation} \label{eq:Throughput_definition}
	R_{\rm a}=\lambda_b p_{\rm a}\left(1-p_{\rm out}\right)R_0,
\end{equation}
where $\lambda_b p_{\rm a}$ is density of active BSs and $R_0\triangleq\log_2\left(1+\hat{\gamma}\right)$. The network throughput can also be regarded as a measure of the area spectral efficiency. In addition to the network throughput, we will also evaluate the user throughput, denoted as $R_{\rm u}$, which is defined as the average throughput per user, given by $R_{\rm u}=\rho p_{\rm a}\left(1-p_{\rm out}\right)R_0$.
In the following sections, we will see that these two throughput metrics will be affected by the BS/user density in different ways, and subsequently, important design guidelines can be drawn.

Energy efficiency is another important performance metric for small cell networks. As BSs consume the largest portion of energy in cellular networks \cite{Hasan11}, we will focus on the total BS power consumption, denoted as $P_{\rm BS}$, to evaluate the network energy efficiency. In practice, the transmit power $P_{\rm t}$ is only one part of the total BS power consumption. To take other power consumption into consideration, we adopt a linear BS power consumption model, which is widely used in the literature and standard organizations \cite{EARTH10} and is given by
\begin{equation}
	P_{\rm BS}=\frac{1}{\eta}P_{\rm t}+MP_{\rm c}+P_{\rm 0},
\end{equation}
where $\eta$ is the power amplifier efficiency, $M$ is the number of transmit antennas, $P_{\rm c}$ accounts for the circuit power of the corresponding RF chain, and $P_{\rm 0}$ is determined by the non-transmission power consumption, including baseband processing, battery backup, cooling, etc.

Taking the BS power model into consideration, the average power consumption per unit area is the transmit power and circuit power consumption from active BSs and the non-transmission power consumption from both active and inactive BSs, which is given by $P_{\rm a}= \lambda_b p_{\rm a} \left(\frac{1}{\eta}P_{\rm t}+MP_{\rm c}\right) +\lambda_b P_{\rm 0}$.
Thus, the network energy efficiency is defined as the ratio of the network throughput to the power consumption per unit area, given by \cite{Quek11}
\begin{equation} \label{eq:EE_definition}
    \eta_{\rm EE}=\frac{R_{\rm a}}{P_{\rm a}}=\frac{p_{\rm a}\left(1-p_{\rm out}\right)R_0} {p_{\rm a}\left(\frac{1}{\eta}P_{\rm t}+MP_{\rm c}\right)+P_{\rm 0}},
\end{equation}
where the unit is bits/J/Hz.

In the following sections, we will first present a simple expression for the outage probability, and then analytically evaluate both the network throughput and energy efficiency.

\section{Outage Probability Analysis} \label{Sec:Outage}

In this section, we will derive a new expression for the outage probability, which is stated in a much simpler form than existing results and will greatly facilitate further performance analysis. To demonstrate the effectiveness of this new expression, we will provide a few key properties of the outage probability, some of which will be used for throughput and energy efficiency analysis.

\subsection{Analysis of the Outage Probability}

As we consider a dense network with a large number of transmitters, it is reasonable to assume an interference-limited scenario \cite{Andrews11}, so the additive noise will be ignored in the following analysis. Later we will justify this assumption through simulation. Then based on Eq. \eqref{eq:SINR_definition}, the outage probability is given by
\begin{equation} \label{eq:Pout_temp1}
	p_{\rm out}= \Pr\left(\frac{P_{\rm t}g_{00}r_0^{-\alpha}}{\sum_{i\in\tilde{\Phi}_b\backslash0}P_{\rm t}g_{i0}R_{i0}^{-\alpha}} \leq \hat{\gamma}\right).
\end{equation}
Since $g_{00}\sim {\rm Gamma} \left(1,M\right)$,  we have
\begin{equation}
	p_{\rm out}= 1- {\rm E}_{r_{0}}\left[{\rm E}_{I}\left[\sum_{n=0}^{M-1}\frac{r_{0}^{\alpha n}}{n!}I^{n}e^{-r_{0}^{\alpha}I}\right]\right],
\end{equation}
where $I\triangleq\hat{\gamma}\sum_{i\in\tilde{\Phi}_b\backslash0} g_{i0}R_{i0}^{-\alpha}$. Denote $s\triangleq r_0^\alpha$, then ${\rm E}_I \left[e^{-sI}\right]$ can be regarded as the Laplace transform of $I$, denoted as $\mathcal{L}_I(s)$. Following the property of the Laplace transform, we have ${\rm E}_I\left[I^n e^{-sI}\right]=\left(-1\right)^n\frac{d^n}{ds^n}\mathcal{L}_I\left(s\right)$, which subsequently gives
\begin{equation} \label{eq:Pout_WithLaplaceForm}
	p_{\rm out} = 1- {\rm E}_{r_{0}}\left[\sum_{n=0}^{M-1}\frac{s^n} {n!}\left(-1\right)^{n}\frac{d^{n}} {ds^n}\mathcal{L}_{I}\left(s\right)\right].
\end{equation}

The major difficulty in the following derivation is how to simplify the $n$th derivative of $\mathcal{L}_{I}\left(s\right)$, which is the common case when dealing with the multi-antenna transmission in the PPP network model. Previous works either use an approximation by Taylor expansion \cite{Hunter08}, or obtain a complicated expression via special functions \cite{Kountouris09}. In \cite{Louie11,Wu12}, the authors derived the closed-form expressions based on the cumulative distribution function (CDF) of the aggregated interference. However, their derivation and results can only deal with the ad-hoc network model, in which the interfering nodes can be arbitrarily close to the considered typical receiver, while in the cellular model, the interfering BSs will be farther away than the home BS. Moreover, their results are still in very complicated forms in terms of some special functions similar to \cite{Kountouris09}. All of these previous results cannot clearly reveal the impacts of BS and user densities and the number of multiple BS antennas.

In contrast to the above approaches, we propose a new method to handle the $n$th derivative of $\mathcal{L}_{I}\left(s\right)$. Specifically, instead of obtaining the complex closed-form expression of the $n$th derivative, we first express it as a recursive form. With the help of linear algebra, the recursive expression can be transformed to a lower triangular Toeplitz matrix form, which possesses nice analytical properties for further performance evaluation. The new expression of the outage probability is given in the following theorem.

\begin{theorem}
\label{Thm:Pout_MatrixForm}
The outage probability in \eqref{eq:Pout_temp1} is given by
    \begin{equation} \label{eq:Pout_MatrixForm2}
    p_{\rm out}= 1-\frac{1}{p_{\rm a}}\left\Vert \left[\left(k_0+\frac{1}{p_{\rm a}}\right)\mathbf{I}-\mathbf{Q}_{M}\right]^{-1}\right\Vert _{1},
    \end{equation}
where $\left\Vert\cdot\right\Vert_{1}$  is the $L_1$ induced matrix norm (i.e., $\left\Vert \mathbf{A}\right\Vert_{1}=\max_{1\leq j \leq n}\sum_{i=1}^{m}\left|a_{ij}\right|$ for $\mathbf{A}\in\mathbb{R}^{m\times n}$), $\mathbf{I}$ is an $M\times M$ identity matrix, $\mathbf{Q}_{M}$ is an $M\times M$ Toeplitz matrix given by
	\begin{equation}
		\mathbf{Q}_{M}=\left[\begin{array}{ccccc}
				0\\
				k_{1} & 0\\
				k_{2} & k_{1} & 0\\
				\vdots &  &  & \ddots\\
				k_{M-1} & k_{M-2} & \cdots & k_{1} & 0
			\end{array}\right], \nonumber
	\end{equation}
$k_{0}=\frac{\frac{2}{\alpha}\hat{\gamma}}{1-\frac{2}{\alpha}} {}_{2}F_{1}\left(1,1-\frac{2}{\alpha};2-\frac{2}{\alpha};-\hat{\gamma}\right)$ and $k_{i}=\frac{\frac{2}{\alpha}\hat{\gamma}^{i}}{i-\frac{2}{\alpha}} {}_{2}F_{1}\left(i+1,i-\frac{2}{\alpha};i+1-\frac{2}{\alpha};-\hat{\gamma}\right)$ for $i\geq 1$, where ${}_{2}F_{1}\left(\cdot\right)$ is the Gauss hypergeometric function.
\end{theorem}

\begin{IEEEproof}
See Appendix \ref{APP:Outage_Derivation}.
\end{IEEEproof}

Compared with the previous results \cite{Hunter08,Louie11,Wu12,Kountouris09}, the expression \eqref{eq:Pout_MatrixForm2} is mathematically more tractable, since we can apply the properties of the Toeplitz matrix \cite{Vecchio03,Commenges84} and the matrix norm for the further analysis. Denote $\mathbf{T}_M\triangleq\frac{1}{p_{\rm a}}\left[\left(k_0+\frac{1}{p_{\rm a}}\right)\mathbf{I}-\mathbf{Q}_{M}\right]^{-1}$, i.e., $p_{\rm out} =1-\left\Vert\mathbf{T}_M\right\Vert _{1}$, then the following lemma provides some basic properties of $\mathbf{T}_M$ to demonstrate the tractability of the outage probability expression.

\begin{lemma} \label{Lemma1}
The matrix $\mathbf{T}_M$ and its $L_1$ norm $\left\Vert \mathbf{T}_{M}\right\Vert _{1}$ have the following properties:
\begin{enumerate}
    \item $\mathbf{T}_M$ is a lower triangular Toeplitz matrix with positive entries, i.e.,
            \begin{equation}
            	\mathbf{T}_{M}=\left[\begin{array}{ccccc}
            			t_0\\
            			t_1 & t_0\\
            			t_2 & t_1 & t_0\\
            			\vdots &  &  & \ddots\\
            			t_{M-1} & t_{M-2} & \cdots & t_1 & t_0
            		\end{array}\right], \nonumber
            \end{equation}
        where $t_n>0$ for $n\in \left[0,M-1\right]$, and the closed-form expression of $t_n$ is given in \eqref{eq:tn_ClosedForm}.
    \item $\frac{\partial\left\Vert \mathbf{T}_{M}\right\Vert _{1}}{\partial p_{{\rm a}}}=\frac{1}{p_{{\rm a}}}\left(\left\Vert \mathbf{T}_{M}^{2}\right\Vert _{1}-\left\Vert \mathbf{T}_{M}\right\Vert _{1}\right)$.
    \item $\left\Vert \mathbf{T}_{M}\right\Vert _{1}$ is bounded as
        \begin{equation} \label{eq:Tm_Bounds}
            \frac{1}{1+p_{\rm a}B_{l}}\leq \left\Vert \mathbf{T}_{M}\right\Vert _{1} \leq\frac{1}{1+p_{\rm a}B_{u}},
        \end{equation}
        where $B_l=k_0-\sum_{i=1}^{M-1}\left(1-\frac{i}{M}\right)k_{i}$ and $B_u=k_0-\sum_{i=1}^{M-1}k_{i}$.
\end{enumerate}
\end{lemma}

\begin{IEEEproof}
See Appendix \ref{APP:Lemma}.
\end{IEEEproof}

Note that $B_l$ and $B_u$ are unrelated to the BS density or the user density, and when $M=1$, $B_l=B_u=k_0$, i.e., \eqref{eq:Tm_Bounds} becomes an identity; while $M\rightarrow\infty$, both $B_l$ and $B_u$ tend to $0$. 

All of these properties will be served as the main tools for performance analysis in the rest of the paper. First, in the following subsection, we will provide some key properties of the outage probability. More detailed investigation of the network throughput and energy efficiency will be pursued in Section \ref{Sec:Throughput}.

\subsection{Key Properties of the Outage Probability} \label{Sec:SubSection_Properties}

Based on Theorem \ref{Thm:Pout_MatrixForm}, we will provide some key properties of the outage probability. In particular, Property \ref{Prop1} and \ref{Prop2} provide insights on the effect of the BS density, while Property \ref{Prop3} and \ref{Prop4} are useful when analyzing how the number of transmit antennas affects system performance.

\begin{property} \label{Prop1}
The outage probability is a decreasing function w.r.t. the BS density, i.e., $\frac{\partial p_{\rm out}}{\partial \lambda_b}\leq 0$, and it is a constant for a given BS-user density ratio $\rho$.
\end{property}

\begin{IEEEproof}
    From  \eqref{eq:Pout_MatrixForm2}, we see that the outage probability $p_{\rm out}$ is a function of the BS density $\lambda_b$ through the BS activity probability $p_{\rm a}$, which is a monotone decreasing function with $\lambda_b$. Therefore, the inequality $\frac{\partial p_{\rm out}}{\partial \lambda_b}\leq 0$ is equivalent with $\frac{\partial p_{\rm out}}{\partial p_{\rm a}}\geq 0$. Based on Lemma \ref{Lemma1}, the derivative of the outage probability with respect to $p_{\rm a}$ is given by
    \begin{equation}
        \frac{\partial p_{{\rm out}}}{\partial p_{{\rm a}}} = -\frac{\partial\left\Vert \mathbf{T}_{M}\right\Vert _{1}}{\partial p_{{\rm a}}} = -\frac{1}{p_{{\rm a}}}\left(\left\Vert \mathbf{T}_{M}^{2}\right\Vert _{1}-\left\Vert \mathbf{T}_{M}\right\Vert _{1}\right).
    \end{equation}
    Since $\left\Vert \mathbf{T}_{M}^{2}\right\Vert _{1} \leq \left\Vert \mathbf{T}_{M}\right\Vert _{1}^{2}$, we have
    \begin{equation}
        \frac{\partial p_{{\rm out}}}{\partial p_{{\rm a}}} = -\frac{1}{p_{{\rm a}}}\left(\left\Vert \mathbf{T}_{M}^{2}\right\Vert _{1}-\left\Vert \mathbf{T}_{M}\right\Vert _{1}\right) \geq -\frac{1}{p_{{\rm a}}}\left(\left\Vert \mathbf{T}_{M}\right\Vert _{1}^{2}-\left\Vert \mathbf{T}_{M}\right\Vert _{1}\right)=\frac{1}{p_{{\rm a}}}\left\Vert \mathbf{T}_{M}\right\Vert _{1}p_{{\rm out}} \geq 0,
    \end{equation}
    which is equivalent to $\frac{\partial p_{\rm out}}{\partial \lambda_b}\leq 0$.
\end{IEEEproof}

This property implies that deploying more BSs will always reduce the outage probability for a typical user. This result actually is not quite straightforward, as increasing the BS density will increase both the signal power and the interference power. An intuitive explanation of this result is that the average received signal power can be shown to scale with the BS density as $\lambda_b^{\frac{\alpha}{2}}$, while the average received aggregate interference power scales as $\left(p_{\rm a}\lambda_b\right)^{\frac{\alpha}{2}}$. For a fixed user density, the BS activity probability $p_{\rm a}$ will decrease as the BS density increases, as shown in \eqref{eq:pa}. Therefore, the interference power increases more slowly than the signal power as the BS density increases, and thus the outage probability decreases.

\begin{remark}
This property is the consequence of considering the explicit BS-user association and the BS activity probability. In previous works, it is assumed that there is always one user for each BS to serve (i.e., $p_{\rm a}=1$), so increasing the BS density will not affect the outage probability \cite{Andrews11}. However, in small cell networks, such as micro-cells and femto-cells, the user density is comparable to the BS density \cite{Andrews12}, so it is necessary to take the BS activity probability into consideration. We will see more results related to the BS/user density in Section \ref{Sec:Throughput}.
\end{remark}

Defining the successful transmission probability as $p_{\rm s}\triangleq 1-p_{\rm out}$, which was called as the coverage probability in \cite{Andrews11}, we can obtain the following property.

\begin{property} \label{Prop2}
The successful transmission probability is bounded by
    \begin{equation} \label{eq:Approximation_ps}
    \frac{1}{1+p_{\rm a}B_{l}}\leq p_{\rm s}\leq\frac{1}{1+p_{\rm a}B_{u}},
    \end{equation}
where $B_l$ and $B_u$ are given in Lemma \ref{Lemma1}.
\end{property}

\begin{IEEEproof}
    Since $p_{\rm s}=\left\Vert \mathbf{T}_{M}\right\Vert _{1}$, this property follows directly from Lemma \ref{Lemma1}.
\end{IEEEproof}

From \eqref{eq:Approximation_ps}, we can obtain an approximation of $p_{\rm s}$, given by
\begin{equation} \label{eq:Approximation_ps_form2}
p_{\rm s} = \left\Vert \mathbf{T}_{M}\right\Vert _{1} \approx \frac{1}{1+p_{\rm a}B},
\end{equation}
where $B$ is between $B_u$ and $B_l$. Note that this approximation separates the effect of the BS density and the effect of the number of BS transmit antennas, as $p_{\rm a}$ is only related to the BS-user density ratio $\rho$, while $B_l$ and $B_u$ are determined by the number of BS antennas, the pathloss exponent and the SINR threshold. Thus it allows us to focus on the impact of the BS density on system performance.

The above properties mainly consider the effect of the BS density. Next we will provide properties showing the effect of multiple BS antennas.

\begin{property} \label{Prop3}
The performance gain of increasing the number of BS antennas from $M$ to $M+1$ in terms of the successful transmission probability is
    \begin{equation} \label{eq:Ps_AntennaGainForm_diff_form}
    p_{\rm s}\left(M+1\right)-p_{\rm s}\left(M\right)=t_M,
    \end{equation}
where $p_{\rm s}\left(M\right)$ denotes the successful transmission probability with $M$ antennas at each BS.
\end{property}

\begin{IEEEproof}
    The result follows from the following equality
    \begin{equation} \label{eq:Ps_AntennaGainForm}
        p_{\rm s}\left(M\right) = \left\Vert \mathbf{T}_{M}\right\Vert _{1}=\sum_{n=0}^{M-1}t_n.
    \end{equation}
\end{IEEEproof}

This property shows the benefit of deploying multiple transmit antennas. First, since $t_n\geq0$, increasing the number of transmit antennas will always increase the successful transmission probability. Second, from \eqref{eq:tn_ClosedForm}, it can be shown that $t_n>t_{n+1}$ for $n\geq0$, which implies that the effect of adding one more antenna on the successful transmission probability diminishes as $n$ increases. Furthermore, when the number of antennas is large, we have the following property:

\begin{figure}
    \begin{center}
    \scalebox{0.7}{\includegraphics{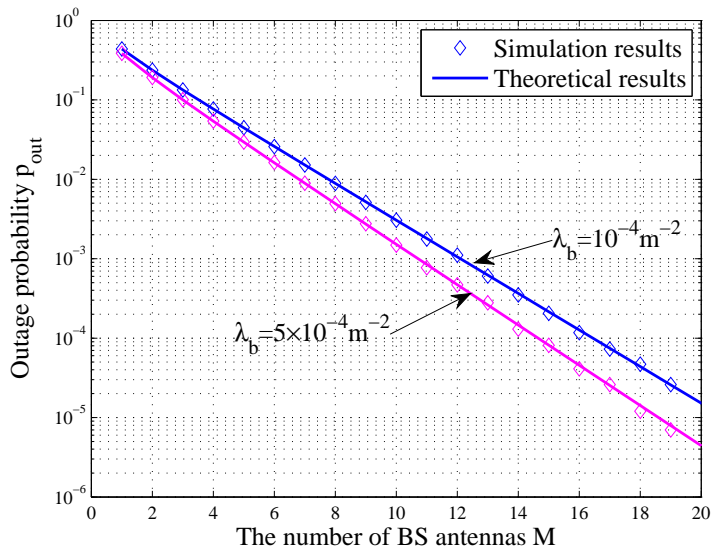}}
    \end{center}
    \caption{The outage probability with different number of transmission antennas with $\alpha=4$, $\hat{\gamma}=1$, $\lambda_u=10^{-3}$ per square meter. The transmit power is $6.3$W, and the noise power considered in the simulation is $\sigma_n^2=-97.5$dBm \cite{Holma09}. }
    \label{fig:pout}
\end{figure}

\begin{property} \label{Prop4}
Denote $p_{\rm out}\left(M\right)$ as the outage probability with $M$ antennas at each BS. We have
    \begin{equation} \label{eq:Prop4_tn}
    \lim_{n\rightarrow\infty}\frac{t_n}{t_{n+1}} = r_u,
    \end{equation}
and
    \begin{equation} \label{eq:Prop4_Pout}
    \lim_{M\rightarrow\infty}\frac{p_{\rm out}\left(M\right)}{p_{\rm out}\left(M+1\right)} = r_u,
    \end{equation}
where $r_u\!\in\!\left(1,1\!+\!\hat{\gamma}^{-1}\right)$ is unrelated to $M$ and is the solution of the equation $\hat{\gamma}^{\frac{2}{\alpha}}\! \int_{\hat{\gamma}^{-\frac{2}{\alpha}}}^{\infty}\! \frac{\left(r_{u}-1\right)dv}{1+v^{\frac{\alpha}{2}}-r_{u}} \!=\! \frac{1}{p_{{\rm a}}}$.
\end{property}

\begin{IEEEproof}
    See Appendix \ref{APP:Prop4}.
\end{IEEEproof}

This property has three implications: 1) It is shown in \eqref{eq:Prop4_tn} that when $M$ is large, the benefit of adding the $\left(n+1\right)$th antenna is $\frac{1}{r_u}$ times smaller than adding the $n$th antenna; 2) Eq. \eqref{eq:Prop4_Pout} implies that when $M$ is large, the outage probability in the logarithmic scale decreases linearly with $M$ with the slope $\log_{10}\left(\frac{1}{r_u}\right)$, which is demonstrated in Fig. \ref{fig:pout}. Moreover, from Fig. \ref{fig:pout}, we see that this linearity holds even for small values of $M$. 3) Since increasing $\lambda_b$ can increase $r_u$, it means that the performance gain of adding one more antenna is greater with a larger BS density. In Fig. \ref{fig:pout}, it is shown that the outage probability decreases faster as $M$ increases for $\lambda_b=5\times10^{-4}$ than $\lambda_b=10^{-4}$ per square meter.

The above discussion demonstrates the effectiveness of the analytical result \eqref{eq:Pout_MatrixForm2} for outage probability. In the next section, we will apply it to evaluate the network throughput and energy efficiency of small cell networks.

\section{Throughput and Energy Efficiency Analyses} \label{Sec:Throughput}

In this section, we will analytically evaluate throughput and energy efficiency of small cell networks, with the main focus on the impact of the BS density and the number of BS antennas.

\subsection{Throughput Analysis}

As shown in Eq. \eqref{eq:Throughput_definition}, the effect of the number of BS antennas on the throughput is the same as that on the successful transmission probability, which has been revealed through Property \ref{Prop3} and \ref{Prop4} in Section \ref{Sec:SubSection_Properties}. In short, the throughput increases as we deploy more BS antennas, but the performance gain diminishes. In the following we will focus on the effect of the BS density.

Substituting \eqref{eq:Pout_MatrixForm2} into \eqref{eq:Throughput_definition}, the network throughput is given by
\begin{equation} \label{eq:R_a_closed_form}
    R_{\rm a}=\lambda_b\left\Vert \left[\left(k_0+\frac{1}{p_{\rm a}}\right)\mathbf{I}-\mathbf{Q}_{M}\right]^{-1}\right\Vert _{1}R_0.
\end{equation}
Then, according to Property \ref{Prop2} of the outage probability, we can get the following lower and upper bounds for the network throughput
\begin{equation} \label{eq:Approximation_Ra}
    \frac{\lambda_b R_0}{\frac{1}{p_{\rm a}}+B_{l}} \leq R_{\rm a} \leq \frac{\lambda_b R_0} {\frac{1}{p_{\rm a}}+B_{u}}.
\end{equation}
Similarly, the user throughput is bounded as
\begin{equation} \label{eq:Approximation_Ru}
    \frac{\rho R_0}{\frac{1}{p_{\rm a}}+B_{l}} \leq R_{\rm u} \leq \frac{\rho R_0}{\frac{1}{p_{\rm a}}+B_{u}}.
\end{equation}
In the following, we will investigate the impact of the BS density on the network and user throughput in two different scenarios.

\paragraph*{Scenario 1 (For a fixed user density)}

We first consider a fixed user density, and will investigate how the network throughput will change with different BS densities. This is of practical relevance, as it corresponds to investigating how much additional gain can be provided if the operator deploys more BSs. In this case, the effect of the BS density on $R_{\rm a}$ is the same as that on $R_{\rm u}$. We will consider the following three different regimes in terms of the BS density.

\begin{figure}
    \begin{center}
    \scalebox{0.7}{\includegraphics{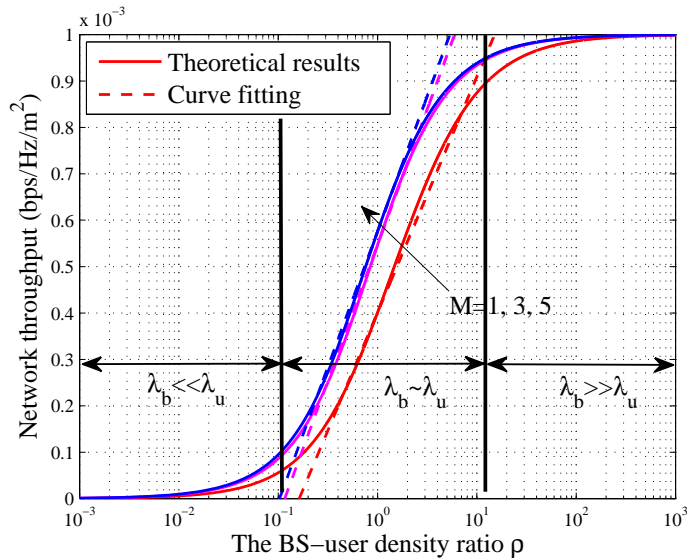}}
    \end{center}
    \caption{The network throughput for different BS-user density ratios, with $\alpha=4$ and $\hat{\gamma}=1$.}
    \label{fig:Ra}
\end{figure}

\begin{itemize}
    \item \textbf{Low BS density regime}: $\lambda_b\ll\lambda_u$ or $\rho\ll 1$, so $p_{\rm a}\approx 1$ (e.g., $p_{\rm a}>0.99$ when $\rho\leq0.1$), which means almost all the BSs are active. From \eqref{eq:R_a_closed_form} we see that in this regime $R_{\rm a}$ increases \emph{linearly} with $\lambda_b$, i.e.,
        \begin{equation}
            R_{{\rm a}}=c_{0}\lambda_{b},
        \end{equation}
        where $c_{0}\triangleq\left\Vert \left[\left(k_{0}+1\right)\mathbf{I} - \mathbf{Q}_{M}\right]^{-1} \right\Vert _{1}R_0$ is unrelated to $\lambda_b$. This is the common case considered in most previous works, such as \cite{Andrews11}.

    \item \textbf{High BS density regime}: $\lambda_b\gg\lambda_u$, so all the users are being served and $p_{\rm a}\approx \frac{1}{\rho}$. From \eqref{eq:Approximation_Ra}, we see that increasing the BS density can increase the throughput, but the improvement is quite limited as
        \begin{equation}
            R_{{\rm a}}\approx\frac{\lambda_{u}R_0}{1+B/\rho}\rightarrow\lambda_{u}R_0\quad{\rm when}\quad\rho\gg1.
        \end{equation}
        So there is no need to further increase $\lambda_b$. This special case is considered in \cite{Lee12}.

    \item \textbf{Medium BS density regime}: $\lambda_b\sim\lambda_u$, i.e., the BS density and the user density are comparable, which is a more practical case for small cell networks \cite{Andrews12}. For this case, the exact expression for the network throughput is given in \eqref{eq:R_a_closed_form}, but it is difficult to get the scaling result in this finite regime. We therefore resort to data fitting, which shows that $R_{\rm a}$ increases \emph{logarithmically} with the BS density, i.e.,
        \begin{equation}
            R_{{\rm a}}\approx c_1\log \rho +c_2,
        \end{equation}
        where $c_1$ and $c_2$ can be determined by data fitting. We have shown some numerical examples to validate such relationship in Fig. \ref{fig:Ra}, while a more accurate characterization will be left to future work.

\end{itemize}

The above analysis shows that when considering explicit BS-user association, the BS-user density ratio is critical to network throughput. This should be carefully taken into consideration when evaluating small cell networks.

\paragraph*{Scenario 2 (With a fixed $\rho$)}

In this case, the BS density varies in proportion to the user density. From \eqref{eq:Approximation_Ru}, we see that $R_{\rm u}$ is the same for a fixed $\rho$; While from \eqref{eq:Approximation_Ra}, we see that $R_{\rm a}$ increases linearly with $\lambda_b$ for a fixed $\rho$. This means that if we keep the BS-user density ratio fixed, the network throughput grows linearly with the BS density while the throughput of a typical user stays the same. Equivalently, this indicates that once the user density increases, the operator can improve the network throughput by deploying more small BSs, while maintaining the QoS for each user, which demonstrates the advantage of small cell networks.

\subsection{Network Energy Efficiency Analysis -- The Effect of the BS Density}

In the following, we will evaluate the network energy efficiency in small cell networks, which will be shown to depend critically on the BS power consumption model. In particular, there is no simple monotonic result with respect to the BS density or multiple BS antennas, and different conclusions will be drawn under different conditions.

By substituting \eqref{eq:Pout_MatrixForm2} into \eqref{eq:EE_definition}, we can obtain the following expression of the energy efficiency
\begin{equation} \label{eq:EE_Expression}
    \eta_{{\rm EE}}=\frac{\left\Vert \left[\left(k_{0}+\frac{1}{p_{{\rm a}}}\right)\mathbf{I}-\mathbf{Q}_{M}\right]^{-1}\right\Vert _{1}R_{0}}{p_{{\rm a}}\left(\frac{1}{\eta}P_{{\rm t}}+MP_{{\rm c}}\right)+P_{0}}.
\end{equation}
Then we can get the following result showing the effect of the BS density.

\begin{proposition}
The energy efficiency is a decreasing function with $\lambda_b$ if
    \begin{equation} \label{eq:P0Inequality}
        \frac{P_{0}}{P_{{\rm BS}}}\geq1-\frac{\left\Vert \left[\left(k_{0}+1\right)\mathbf{I}-\mathbf{Q}_{M}\right]^{-2}\right\Vert _{1}}{\left\Vert \left[\left(k_{0}+1\right)\mathbf{I}-\mathbf{Q}_{M}\right]^{-1}\right\Vert _{1}}\triangleq\gamma_{P_{0}}.
    \end{equation}
Otherwise, the energy efficiency first increases and then decreases as $\lambda_b$ increases, and there is a non-zero optimal BS density  $\lambda_b^*$ that maximizes the energy efficiency. The approximated maximum energy efficiency is $\eta_{{\rm EE}}^{*}\approx\frac{R_0}{\left(\sqrt{\frac{1}{\eta}P_{{\rm t}}+MP_{{\rm c}}}+\sqrt{P_{0}B}\right)}$ and the corresponding optimal BS density is
    \begin{equation} \label{eq:Optimal_BSdensity}
        \lambda_{b}^{*} \approx \frac{1}{\mu}\left[1-\left(1- \sqrt{\frac{P_{0}}{B\left(\frac{1}{\eta}P_{{\rm t}}+MP_{{\rm c}}\right)}}\right)^{-\frac{1}{\mu}}\right] \lambda_{u}.
    \end{equation}
\end{proposition}

\begin{IEEEproof}
    Since $p_{\rm a}$ is a monotone decreasing function with $\lambda_b$, the effect of the BS density $\lambda_b$ on the energy efficiency is the opposite as that of the BS activity probability $p_{\rm a}$. Therefore, the condition \eqref{eq:P0Inequality} is derived by investigating the derivative of \eqref{eq:EE_Expression} w.r.t. $p_{\rm a}$, while the approximated optimal energy efficiency and the corresponding BS density can be obtained through \eqref{eq:Approximation_ps_form2}.
\end{IEEEproof}

From this result, we can see that the non-transmission power consumption $P_0$ plays a critical role in the energy efficiency. Particularly, when $\frac{P_{0}}{P_{{\rm BS}}}\geq \gamma_{P_0}$, increasing the BS density will always decrease energy efficiency, although it can improve the throughput. On the other hand, when $\frac{P_0}{P_{\rm BS}}< \gamma_{P_0}$, there is a non-zero BS density that can achieve the maximum energy efficiency, which is instructive when designing and operating a cellular network.

\subsection{Network Energy Efficiency Analysis -- The Effect of the Number of BS Antennas}

We have seen that increasing the number of transmit antennas will increase the throughput, but it will also consume more circuit power $P_{\rm c}$. We will next investigate how the BS antenna number will affect the overall network energy efficiency. Denote $\eta_{\rm EE}\left(M\right)$ as the energy efficiency with $M$ antennas per BS, then by substituting \eqref{eq:Ps_AntennaGainForm} in \eqref{eq:EE_definition}, the energy efficiency can be written as
\begin{equation} \label{eq:EE_Sum_tn_Form}
    \eta_{{\rm EE}}\left(M\right) = \frac{\sum_{n=0}^{M-1}t_n}{\frac{1}{\eta}P_{\rm t}+MP_{\rm c}+ \frac{P_0}{p_{\rm a}}}R_0.
\end{equation}
Then the effect of the number of BS antennas on the energy efficiency is given in the following proposition.

\begin{proposition} \label{Propo2}
    There is an optimal number of BS transmit antennas $M^*$ that maximizes the energy efficiency. When $M>M^*$, increasing $M$ will decrease the energy efficiency, while for $M<M^*$, deploying more antennas can improve the energy efficiency. The optimal $M^*$ is the greatest integer that is smaller than the solution of the equation
    \begin{equation} \label{eq:Optimal_M_equality}
        F\left(M\right)=\frac{p_{{\rm a}}\left(\frac{1}{\eta}P_{{\rm t}}\right)+P_{0}}{p_{{\rm a}}P_{{\rm c}}},
    \end{equation}
    where $F\left(M\right)\triangleq\frac{p_{\rm s}\left(M\right)}{t_{M-1}}-M$.
\end{proposition}

\begin{IEEEproof}
See Appendix \ref{APP:Propo2}.
\end{IEEEproof}

Since $F\left(M\right)=\frac{1}{t_{M-1}}\sum_{n=0}^{M-2}\left(t_{n}-t_{M-1}\right)$, it is obvious that $F\left(M\right)$ is an increasing function with $M$. Then if we could deploy BSs with a smaller $P_{\rm c}$, then the optimal number of transmit antennas would be larger. Subsequently, both the spectral efficiency and the energy efficiency can be improved.

An extreme case is $M^*=1$, which implies that using single-antenna BSs can provide higher energy efficiency than using the multi-antenna BSs. For this case, we can find the condition from \eqref{eq:Optimal_M_Inequality} as $P_{{\rm c}}\geq\frac{k_{1}\left(p_{{\rm a}}\frac{1}{\eta}P_{{\rm t}}+P_{0}\right)}{1+\left(k_{0}-k_{1}\right)p_{\rm a}}$, where the right hand side of this inequality is a monotone function w.r.t. $p_{\rm a}$, which means if the condition
\begin{equation} \label{eq:Condition_SingleAntenna}
P_{{\rm c}}\geq\max\left(k_1P_0, \frac{k_{1}\left(\frac{1}{\eta}P_{{\rm t}}+P_{0}\right)}{1+k_{0}-k_{1}} \right)\triangleq\gamma_{P_{\rm c}}
\end{equation}
is satisfied, for any BS and user densities, deploying single-antenna BSs is more energy efficient than multi-antenna BSs. Therefore, multi-antenna BSs are preferable in terms of energy efficiency only when the circuit power consumption is smaller than the threshold $\gamma_{P_{\rm c}}$

To summarize, Table \ref{tab:Throughput_EE} shows the main results on the effect of the BS density and the number of transmit antennas on the network throughput and energy efficiency.

\begin{table}
    \centering\small
    \caption{\label{tab:Throughput_EE}The effect of $\lambda_b$ and $M$ on $R_{\rm a}$ and $\eta_{\rm EE}$}
\begin{tabular}{|c|c|c|}
    \hline
                    & Network throughput & Energy efficiency \\ \hline
     \multirow{2}{*}{$\lambda_b$}    & ${\rm for}\;{\rm fixed}\;\lambda_{u}:\;\begin{cases}
    R_{{\rm a}}=c_{0}\lambda_{b} & {\rm for}\;\lambda_{b}\ll\lambda_{u}\\
    R_{{\rm a}}\approx c_{1}\log\lambda_{b}+c_{2} & {\rm for}\;\lambda_{b}\sim\lambda_{u}\\
    R_{a}\rightarrow\lambda_{u}r & {\rm for}\;\lambda_{b}\gg\lambda_{u}
    \end{cases}$ & \multirow{2}{*}{\parbox{0.4\textwidth}{\begin{itemize} \item If $\frac{P_0}{P_{\rm BS}}\geq \gamma_{P_0}$, $\eta_{\rm EE}$ decreases with $\lambda_b$; \item Otherwise, there is one $\lambda_b^*$ to maximize $\eta_{\rm EE}$ \end{itemize}}} \\
    & for fixed $\rho$: $R_{\rm u}$ is fixed, while $R_{\rm a}$ is linear w.r.t. $\lambda_b$ &  \\ \hline
    $M$ & $R_{{\rm a}}=\lambda_{b}p_{{\rm a}}r\sum_{n=0}^{M-1}t_{n}$ where $\lim_{n\rightarrow\infty}\frac{t_{n}}{t_{n+1}}=r_{u}$ & \parbox{0.4\textwidth}{\begin{itemize} \item If $P_{\rm c}\geq \gamma_{P_{\rm c}}$, single-antenna BSs achieve the maximum $\eta_{\rm EE}$; \item Otherwise, there is one $M^*>1$ to maximize $\eta_{\rm EE}$ \end{itemize}} \\ \hline
\end{tabular}

\end{table}

\section{Numerical Results} \label{Sec:NumResults}
In this section, we will demonstrate our results through simulation. An additive noise is considered in all the simulations, to test the interference-limited assumption. Furthermore, we will also run simulations in a regular grid-based network model, to show that our conclusions drawn from the random network model hold in general. The pathloss exponent is $\alpha=4$, and the SINR threshold is set to $1$.

\begin{figure}
    \begin{center}
    \scalebox{0.7}{\includegraphics{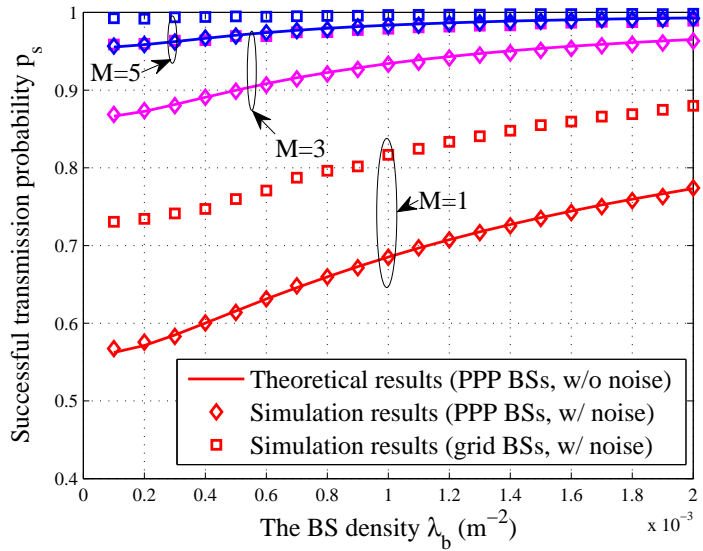}}
    \end{center}
    \caption{The successful transmission probability vs. the BS density for different number of BS antenna, with $\alpha=4$, $\hat{\gamma}=1$, $\lambda_u=10^{-3}{\rm m}^{-2}$. The transmit power is $6.3$W, and noise power considered in the simulation is $\sigma_n^2=-97.5$dBm \cite{Holma09}.}
    \label{fig:pout_BSDensity}
\end{figure}

Fig. \ref{fig:pout_BSDensity} lists the successful transmission probability with different BS densities and different numbers of BS antennas, where the user density is $\lambda_u=10^{-3}$ per square meter. We see that increasing the BS density, or increasing the number of transmit antennas can increase the successful transmission probability. In particular, there is a significant gain from $M=1$ to $M=3$, while the gain becomes smaller from $M=3$ to $M=5$. It is also shown that the numerical results based on \eqref{eq:Pout_MatrixForm2} fits the simulation results, which means that the influence of the additive noise is negligible and our approximation of $\tilde{\Phi}_b$ is accurate. To confirm our conclusions based on the random network model, we also simulate a grid-based model with the same BS density, where each cell is modeled as a hexagon. From Fig. \ref{fig:pout_BSDensity}, we can find that the performance of the hexagonal-cell network provides an upper bound compared to the random network model, which was also shown and explained in \cite{Andrews11}, but both network models have the same trend, i.e., increasing the BS density or the BS antenna number will increase the successful transmission probability.

\begin{figure}
    \begin{center}
    \scalebox{0.7}{\includegraphics{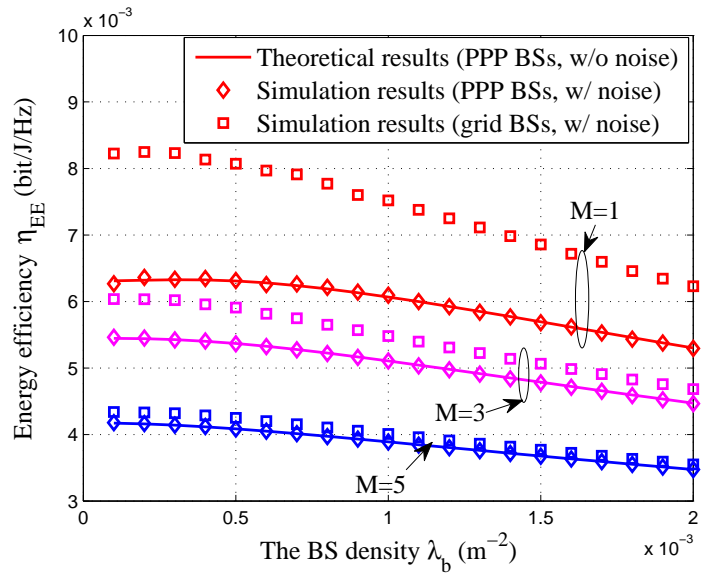}}
    \end{center}
    \caption{Energy efficiency vs. BS density for different number of BS antenna with $\alpha=4$, $\hat{\gamma}=1$, $\lambda_u=10^{-3}{\rm m}^{-2}$, $\eta=0.32$, $P_{\rm t}=6.3$W, $P_{\rm c}=35$W, $P_0=34$W, and the noise power considered in simulation is $\sigma_n^2=-97.5$dBm. In the figure, it is shown that the optimal BS density for $M=1$ case is about $0.3\times10^{-3}{\rm m}^{-2}$ for both the random and grid-based models.}
    \label{fig:EE}
\end{figure}

Fig. \ref{fig:EE} shows the change of the network energy efficiency as the BS density increases. The system setting is the same with Fig. \ref{fig:pout_BSDensity}. For the power consumption model, we consider a micro-BS with $\eta=0.32$, $P_{\rm t}=6.3$W, $P_{\rm c}=35$W, and $P_0=34$W \cite{EARTH10}. By substituting these values, we can find that the condition \eqref{eq:Condition_SingleAntenna} is satisfied, which means it is more energy efficient to deploy single-antenna BSs. Moreover, for $M=1$, there is a non-zero optimal BS density, which can be calculated from \eqref{eq:Optimal_BSdensity} as $\lambda_b^*=0.32\lambda_u$. On the other hand, when $M>2$, the energy efficiency is a decreasing function w.r.t. $\lambda_b$. These results are confirmed by the simulation results in Fig. \ref{fig:EE}. In addition, we can find that compared with the random network model, the performance of the hexagonal-cell network still provides an upper bound, while the trends of both network models are the same. Even the optimal BS density $\lambda_b^*$ for a grid-based model is close to our analytical result. Interestingly, the analytical result for the random network model gets closer to the grid-based model as $M$ increases.

\section{Conclusions} \label{Sec:Conclusions}

In this paper, we developed a new set of analytical results for performance analysis in a random cellular network with multi-antenna BSs. Based on these new results, we investigated the effect of the BS density and number of transmit antennas on the network throughput and energy efficiency. In particular, we characterized the scaling of the network throughput with respect to the BS density and the number of BS antennas, respectively. Moreover, we found that unlike the network throughput, increasing the BS density or the number of BS antennas can increase the energy efficiency only when the different components of BS power consumption satisfy certain conditions. Otherwise, the energy efficiency will always decrease.

The results derived in this paper are particularly useful for analyzing random spatial networks with multi-antenna transmission. Future research directions would include extending the results to other multi-antenna transmission techniques, such as multi-user MIMO or interference cancellation. It is also interesting to investigate interference management techniques such as network MIMO, including its effect on energy efficiency.

\appendix
\subsection{Proof of Theorem \ref{Thm:Pout_MatrixForm}} \label{APP:Outage_Derivation}

Denote $x_{n}=\frac{s^{n}\left(-1\right)^{n}}{n!}\mathcal{L_{I}}^{\left(n\right)}\left(s\right)$, where $\mathcal{L_{I}}^{\left(n\right)}\left(s\right)$ is the $n$th derivative of $\mathcal{L_{I}}\left(s\right)$, then the outage probability in \eqref{eq:Pout_WithLaplaceForm} can be expressed as
\begin{equation} \label{eq:Pout_Sum_Antnna_Gain_Appendix}
	p_{\rm out}=1-\sum_{n=0}^{M-1}{\rm E}_{r_{0}}\left[x_n\right]\quad\textrm{for }M\geq1.
\end{equation}

Then the main objective is to derive an explicit expression for $x_n$. We start from the Laplace transform of $I$, given by
\begin{equation} \label{eq:LaplaceTransform_Orginal}
	\mathcal{L}_{I}\left(s\right) = {\rm E}_{I}\left[\exp{\left(-s\hat{\gamma}\sum_{i\in\tilde{\Phi}_b\backslash0} g_{i0}R_{i0}^{-\alpha}\right)}\right].
\end{equation}
As $g_{i0}\sim {\rm Exp}\left(1\right)$ are independent for different $i$, $\mathcal{L}_{I}\left(s\right)$ can be expressed as
\begin{equation}
	\mathcal{L}_{I}\left(s\right) = {\rm E}_{\tilde{\Phi}_{b}} \left[\prod_{i\in\tilde{\Phi}_{b}\backslash0} {\rm E}_{g_{i0}} \left[\exp\left(-s\hat{\gamma}g_{i0}R_{i0}^{-\alpha}\right) \right] \right] = {\rm E}_{\tilde{\Phi}_{b}} \left[\prod_{i\in\tilde{\Phi}_{b}\backslash0} \frac{1}{1+s\hat{\gamma}R_{i0}^{-\alpha}} \right].
\end{equation}
Using the probability generating functional (PGFL) of PPP \cite{Andrews11}, $\mathcal{L}_I \left( s \right)$ can be expressed as
\begin{equation} \label{eq:LaplaceTransform_beforeDerivative}
	\mathcal{L}_{I}\left(s\right)= \exp{\left[-\pi\lambda_bp_{\rm a} \intop_{r_0^2}^{\infty}\left(1-\frac{1}{1+s\hat{\gamma}u^{-\frac{\alpha}{2}}}\right)du\right]}.
\end{equation}
Then, the $n$th derivative of $\mathcal{L}_I\left(s\right)$ w.r.t. $s$ can be written as the following recursive form
\begin{eqnarray} \label{eq:LaplaceTransmform_Derivative}
	\mathcal{L}_{I}^{\left(n\right)}\left(s\right) = \pi\lambda_bp_{\rm a}\sum_{i=0}^{n-1}\!\left( \!\!\!\begin{array}{c}
\!n-1\!\\
\!i\!
\end{array} \!\!\!\right)\!\left(-1\right)^{n-i}\left(n-i\right)!s^{\frac{2}{\alpha}-n+i} \times  \hat{\gamma}^{\frac{2}{\alpha}} \!\! \intop_{\hat{\gamma}^{-\frac{2}{\alpha}}}^{\infty} \!\! \frac{\left(v^{-\frac{\alpha}{2}}\right)^{n-i}dv}{\left(1+v^{-\frac{\alpha}{2}}\right)^{n-i+1}} \times\mathcal{L_{I}}^{\left(i\right)}\left(s\right).
\end{eqnarray}

By substituting $s\!=\!r_0^\alpha$ into \eqref{eq:LaplaceTransform_beforeDerivative}, we have   $x_{0}\!=\!\mathcal{L}_{I}\left(s\right)\!=\!\exp{\!\left(-\pi\lambda_bp_{\rm a} k_{0}r_{0}^{2}\right)}$, where $k_0$ is given by $k_{0}= \hat{\gamma}^{\frac{2}{\alpha}} \intop_{\hat{\gamma}^{-\frac{2}{\alpha}}}^{\infty} \frac{1}{1+v^{\frac{\alpha}{2}}}dv$.
From \eqref{eq:LaplaceTransmform_Derivative}, we get for $n\geq1$,
\begin{equation} \label{eq:xt_RecurrenceForm}
	x_n=\pi\lambda_bp_{\rm a} r_{0}^{2}\sum_{i=0}^{n-1}\frac{n-i}{n}k_{n-i}x_{i},
\end{equation}
where $k_{i}=\hat{\gamma}^{\frac{2}{\alpha}} \intop_{\hat{\gamma}^{-\frac{2}{\alpha}}}^{\infty} \frac{1}{\left(1+v^{\frac{\alpha}{2}}\right)^{i} \left(1+v^{-\frac{\alpha}{2}}\right)} dv$
for $i\geq1$. Note that $k_0$ and $k_i$ can be expressed as the Gauss hypergeometric function, as shown in the statement of Theorem \ref{Thm:Pout_MatrixForm}.

By now, we have obtained a linear recurrence relation of $x_t$, which will be solved in explicit expression via linear algebra. Denote $\mathbf{x}_{M}=\left[x_{1},x_{2},\ldots,x_{M}\right]^{T}$, $\mathbf{k}_{M}=\left[k_{1},k_{2},\ldots,k_{M}\right]^{T}$ and
\begin{equation}
\mathbf{G}_{M}=\left[\begin{array}{ccccc}
0\\
\frac{1}{2}k_{1} & 0\\
\frac{2}{3}k_{2} & \frac{1}{3}k_{1} & 0\\
\vdots &  & \ddots & 0\\
\frac{M-1}{M}k_{M-1} & \frac{M-2}{M}k_{M-2} & \cdots & \frac{1}{M}k_{1} & 0
\end{array}\right], \nonumber
\end{equation}
then Eq. \eqref{eq:xt_RecurrenceForm} can be represented in a matrix form as $\mathbf{x}_{M}=ax_{0}\mathbf{k}_{M}+a\mathbf{G}_{M}\mathbf{x}_{M}$ where $a=\pi\lambda_bp_{\rm a} r_{0}^{2}$.
Since $\mathbf{G}_M$ is a strictly lower triangular matrix, we have $\mathbf{G}_M^n=0$ for $n\geq M$. Based on this property, after iterating, $\mathbf{x}_M$ can be written as
\begin{equation}
	\mathbf{x}_{M}=\sum_{n=1}^{M}a^{n}x_{0}\mathbf{G}_{M}^{n-1}\mathbf{k}_{M},
\end{equation}
which already gives a closed-form expression for the outage probability as in \eqref{eq:Pout_Sum_Antnna_Gain_Appendix}. In the following, we will further simplify this expression.

Define
\begin{equation}
\mathbf{Q}_{M+1}\triangleq\left[\begin{array}{ccccc}
0\\
k_{1} & 0\\
k_{2} & k_{1} & 0\\
\vdots &  &  & \ddots\\
k_{M} & k_{M-1} & \cdots & k_{1} & 0
\end{array}\right], \nonumber
\end{equation}
then it can be proved that the following equality 
\begin{equation}
    \mathbf{G}_{M}^{n-1}\mathbf{k}_{M}= \frac{1}{n!}\mathbf{Q}_{M+1}^{n}\left(2:M+1,1\right) \quad\textrm{for }n\in\mathbb{N}^{+}
\end{equation}
holds for $M\geq 1$, where $\mathbf{Q}_{M+1}^{n}\left(2:M+1,1\right)$ represents the elements from the second to the $\left(M+1\right)$th row in the first column of $\mathbf{Q}_{M+1}$.

Therefore, $\mathbf{x}_{M}$ can be written as $\mathbf{x}_{M}=x_0\sum_{n=1}^{M}\frac{1}{n!}a^{n}\mathbf{Q}_{M+1}^{n}\left(2:M+1,1\right)$.
By substituting $\mathbf{x}_{M}$ to \eqref{eq:Pout_Sum_Antnna_Gain_Appendix} and using the $L_1$ induced matrix norm, the outage probability is given by $p_{\rm out}=1-{\rm E}_{r_0}\left[x_0+\left\Vert\mathbf{x}_{M-1}\right\Vert_1\right]$,
which is equivalent to
\begin{equation} \label{eq:Pout_general_MatrixExp_Appendix}
    p_{\rm out}=1-{\rm E}_{r_0}\left[\left\Vert x_0\sum_{n=0}^{M-1}\frac{1}{n!}a^n\mathbf{Q}_M^n \right\Vert_1\right].
\end{equation}

Lastly, since $r_0$ is the distance between a typical user to its nearest BS, using the null probability of a PPP, the complementary CDF of $r_0$ is \cite{Andrews11}
\begin{equation}
	\bar{F}\left(r_0\right)=\Pr\left(\textrm{No BS is in the area }\pi r_0^2\right)=e^{-\pi\lambda_{b}r_{0}^{2}}.
\end{equation}
Then, after taking expectation w.r.t. $r_0$ in \eqref{eq:Pout_general_MatrixExp_Appendix}, the outage probability is given by
\begin{equation}
	p_{\rm out}= 1- \frac{1}{1+k_{0}p_{\rm a}} \sum_{n=0}^{M-1}\left(\frac{p_{\rm a}}{1+k_{0}p_{\rm a}}\right)^{n}\left\Vert \mathbf{Q}_{M}^{n}\right\Vert _{1},
\end{equation}
which is equivalent to \eqref{eq:Pout_MatrixForm2} by Taylor expansion.

\subsection{Proof of Lemma \ref{Lemma1}} \label{APP:Lemma}
Firstly, since $\mathbf{Q}_M$ is a lower triangular Toeplitz matrix, $\mathbf{T}_M$ is also a lower triangular Toeplitz matrix \cite{Vecchio03,Commenges84}, and the recurrence formula of $t_n$ is given by \cite{Vecchio03,Commenges84} as $t_n=c\sum_{i=0}^{n-1}k_{n-i}t_i$,
where $t_0=\frac{1}{1+k_0p_{\rm a}}$ and $c=\frac{p_{\rm a}}{1+k_0p_{\rm a}}$. Furthermore, the closed-form expression of $t_n$ ($n\geq1$) can be derived based on \cite{Vecchio03}, which is given as
\begin{equation} \label{eq:tn_ClosedForm}
    t_{n}=\frac{1}{1+k_{0}p_{{\rm a}}} \sum_{i=1}^{n}c^{i} \sum_{\begin{array}{c}
    i_{1}+\cdots+i_{n}=i\\
    i_{1}+\cdots+ni_{n}=n
    \end{array}}\left(\begin{array}{c}
    i\\
    i_{1},\cdots,i_{n}
    \end{array} \right)\left(\frac{k_{1}}{k_{0}}\right)^{i_{1}} \cdots\left(\frac{k_{n}}{k_{0}}\right)^{i_{n}}.
\end{equation}

Secondly, to prove the rest of the lemma, we define $\mathbf{A}\triangleq\left(k_{0}+\frac{1}{p_{\rm a}}\right)\mathbf{I}-\mathbf{Q}_{M}$. Then the derivative of $\left\Vert \mathbf{T}_{M}\right\Vert _{1}$ w.r.t. $p_{\rm a}$ is given by
\begin{equation}
    \frac{\partial\left\Vert \mathbf{T}_{M}\right\Vert _{1}}{\partial p_{{\rm a}}} = \left\Vert \frac{\partial\mathbf{T}_{M}{}_{1}}{\partial p_{{\rm a}}}\right\Vert _{1}=\left\Vert \frac{p_{{\rm a}}\frac{\partial\mathbf{A}^{-1}}{\partial p_{{\rm a}}}-\mathbf{A}^{-1}}{p_{{\rm a}}^{2}}\right\Vert _{1},
\end{equation}
Since $\frac{\partial\mathbf{A}^{-1}}{\partial p_{{\rm a}}}=-\mathbf{A}^{-1}\frac{\partial\mathbf{A}}{\partial p_{{\rm a}}}\mathbf{A}^{-1}$,
we can obtain $\frac{\partial\left\Vert \mathbf{T}_{M}\right\Vert _{1}}{\partial p_{{\rm a}}} = \frac{1}{p_{{\rm a}}}\left(\left\Vert \mathbf{T}_{M}^{2}\right\Vert _{1}-\left\Vert \mathbf{T}_{M}\right\Vert _{1}\right)$.

Thirdly, to derive the upper bound of $\left\Vert \mathbf{T}_{M}\right\Vert _{1}$, we rewrite $\left\Vert \mathbf{T}_{M}\right\Vert _{1}=\frac{1}{p_{{\rm a}}}c\left\Vert \left(\mathbf{I}-c\mathbf{Q}_{M}\right)^{-1}\right\Vert _{1}$, where $c=\frac{p_{\rm a}}{1+k_0p_{\rm a}}$. Since $\left(\mathbf{I}-c\mathbf{Q}_{M}\right)\left(\mathbf{I}-c\mathbf{Q}_{M}\right)^{-1}=\mathbf{I}$, we have
\begin{equation}
\left(\mathbf{I}-c\mathbf{Q}_{M}\right)^{-1} = \mathbf{I}+c\mathbf{Q}_M\left(\mathbf{I}-c\mathbf{Q}_{M}\right)^{-1}.
\end{equation}
Then, using the triangle inequality, we can obtain
\begin{equation}
    \left\Vert \left[\mathbf{I}-c\mathbf{Q}_{M}\right]^{-1}\right\Vert_{1}\leq \left\Vert\mathbf{I}\right\Vert_{1}+c\left\Vert \mathbf{Q}_M\right\Vert_{1}\left\Vert \left[\mathbf{I}-c\mathbf{Q}_{M}\right]^{-1}\right\Vert_{1},
\end{equation}
which can be written as $\left\Vert \left[\mathbf{I}-c\mathbf{Q}_{M}\right]^{-1}\right\Vert_{1}\leq \frac{\left\Vert\mathbf{I}\right\Vert_{1}} {1-c\left\Vert \mathbf{Q}_M\right\Vert_{1}}$.

As $\left\Vert\mathbf{I}\right\Vert_{1}=1$, and $\left\Vert \mathbf{Q}_M\right\Vert_{1}=\sum_{i=1}^{M-1}k_i$, we can get an upper bound of $\left\Vert \mathbf{T}_{M}\right\Vert _{1}$ as
\begin{equation}
    \left\Vert \mathbf{T}_{M}\right\Vert _{1} \leq \frac{1}{\left(1+k_0 p_{\rm a}\right)\left(1-c\sum_{i=1}^{M-1}k_i\right)} = \frac{1}{1+p_{\rm a}\left(k_0-\sum_{i=1}^{M-1}k_i\right)}.
\end{equation}

For the lower bound, we define $\mathbf{x}\triangleq\left[1,1,\ldots,1\right]^{T}$ and $\mathbf{y}\triangleq\mathbf{A}\mathbf{x}$. As $\mathbf{A}$ is a nonsingular matrix, then we have $\mathbf{x}=\mathbf{A}^{-1}\mathbf{y}$. Using the inequality $\left\Vert \mathbf{x}\right\Vert _{1}\leq\left\Vert \mathbf{A}^{-1}\right\Vert _{1}\left\Vert \mathbf{y}\right\Vert _{1}$, we get $\left\Vert \mathbf{A}^{-1}\right\Vert _{1}\geq \frac{\left\Vert \mathbf{x}\right\Vert_{1}}{\left\Vert \mathbf{y}\right\Vert_{1}}$. Since $\mathbf{y}=\mathbf{A}\mathbf{x}$, we can get $\left\Vert \mathbf{y}\right\Vert_{1}=M\left(k_{0}+\frac{1}{p_{\rm a}}\right)-\left(M-1\right)k_1-\cdots-k_{M-1}$. Therefore, we have the following lower bound of $\left\Vert \mathbf{T}_{M}\right\Vert _{1}$
\begin{equation}
    \left\Vert \mathbf{T}_{M}\right\Vert _{1}\geq \frac{1}{p_{\rm a}}\frac{\left\Vert \mathbf{x}\right\Vert_{1}}{\left\Vert \mathbf{y}\right\Vert_{1}} = \frac{1}{1+p_{\rm a}\left(k_0-\sum_{i=1}^{M-1}k_i+\sum_{i=1}^{M-1}\frac{i}{M}k_i\right)}.
\end{equation}

Note that 
it can be shown that $k_i>k_{i+1}$ for $i\in\mathbb{N}$, and $\sum_{i=1}^{\infty}k_i=k_0$. Therefore, both $B_l=k_0-\sum_{i=1}^{M-1}\left(1-\frac{i}{M}\right)k_{i}$ and $B_u=k_0-\sum_{i=1}^{M-1}k_{i}$ are positive and the gap $B_l-B_u=\sum_{i=1}^{M-1}\frac{i}{M}k_i$ will be a decrease function with $M$ when $M$ is larger than a certain value.

\subsection{Proof of Property \ref{Prop4} of the Outage Probability} \label{APP:Prop4}
Define the power series $F\left(u\right)=\sum_{n=0}^{\infty}t_nu^n$. By substituting \eqref{eq:tn_ClosedForm}, $F\left(u\right)$ can be written as $F\left(u\right) = \frac{1}{1+k_0p_{\rm a}}\frac{1}{1-\frac{k_0p_{\rm a}}{1+k_0p_{\rm a}}G\left(u\right)}$, 
where $G\left(u\right)\!=\!\sum_{i=1}^{\infty}\frac{k_i}{k_0}u^i$. The radius of convergence of $G\left(u\right)$ is $1\!+\!\hat{\gamma}^{-1}$, as $\lim_{i\rightarrow\infty}\frac{k_i}{k_{i+1}}\!=\!1\!+\!\hat{\gamma}^{-1}$. Therefore, the radius of convergence of $F\left(u\right)$ will be the solution $r_u$ of the equation
\begin{equation}
1-\frac{k_0p_{\rm a}}{1+k_0p_{\rm a}}G\left(r_u\right)=0,
\end{equation}
for $r_u\in\left(1,1+\hat{\gamma}^{-1}\right)$. Note that $r_u\rightarrow 1+\hat{\gamma}^{-1}$ if and only if the BS activity probability $p_{\rm a}\rightarrow0$, due to $G\left(1+\hat{\gamma}^{-1}\right)\rightarrow \infty$. Therefore, we have Eq. \eqref{eq:Prop4_tn}.

Next, since the outage probability is $p_{\rm out}=1-\sum_{n=0}^{M-1}t_n$, then
\begin{equation}
\lim_{M\rightarrow\infty}\frac{p_{\rm out}\left(M+1\right)}{p_{\rm out}\left(M\right)} = 1-\lim_{M\rightarrow\infty}\frac{t_M}{1-\sum_{n=0}^{M-1}t_n}= 1-\lim_{M\rightarrow\infty}\frac{1}{\sum_{n=M}^{\infty}\frac{t_n}{t_M}}.
\end{equation}
Based on Eq. \eqref{eq:Prop4_tn}, the above equation can be written as
\begin{equation}
\lim_{M\rightarrow\infty}\frac{p_{\rm out}\left(M+1\right)}{p_{\rm out}\left(M\right)} = 1-\lim_{M\rightarrow\infty}\frac{1}{\sum_{n=0}^{\infty}\left(\frac{1}{r_u}\right)^n} = \frac{1}{r_u}.
\end{equation}

\subsection{Proof of Proposition \ref{Propo2} of the Energy Efficiency} \label{APP:Propo2}

From \eqref{eq:EE_Sum_tn_Form}, we can find that it is not possible that the two inequalities $\eta_{{\rm EE}}\left(M\right)\leq\eta_{{\rm EE}}\left(M-1\right)$ and $\eta_{{\rm EE}}\left(M\right)\leq\eta_{{\rm EE}}\left(M+1\right)$
hold simultaneously, which implies that it would never happen that the energy efficiency first decreases and then increases as we keep increasing $M$. Moreover, we have $\lim_{M\rightarrow\infty}\eta_{{\rm EE}}\left(M\right)=0$, and $\eta_{{\rm EE}}\left(1\right)>0$.
Considering all these facts, there can only be two different cases for the effect of $M$ on the energy efficiency: 1) Energy efficiency decreases with $M$, so deploying a single antenna at each BS is more energy efficient than using multiple antennas; 2) Deploying multi-antenna BSs can achieve higher energy efficiency than single-antenna BSs and there is an optimal value of $M$.

To determine the optimal number of transmit antennas $M^*$, we consider the inequalities
\begin{equation}
\begin{cases}
\eta_{{\rm EE}}\left(M^*\right)\geq\eta_{{\rm EE}}\left(M^*-1\right)\\
\eta_{{\rm EE}}\left(M^*\right)\geq\eta_{{\rm EE}}\left(M^*+1\right).
\end{cases} \nonumber
\end{equation}
Substituting \eqref{eq:EE_Sum_tn_Form} to the above inequalities, we can find that the optimal $M^*$ satisfies the condition
\begin{equation} \label{eq:Optimal_M_Inequality}
\frac{\sum_{n=0}^{M^*-1}t_{n}}{t_{M^*-1}}-M^* \leq \frac{p_{{\rm a}}\left(\frac{1}{\eta}P_{{\rm t}}\right)+P_{0}}{p_{{\rm a}}P_{{\rm c}}} \leq \frac{\sum_{n=0}^{M^*}t_{n}}{t_{M^*}}-\left(M^*+1\right).
\end{equation}
Define the function $F\left(M\right)\triangleq\frac{p_{\rm s}\left(M\right)}{t_{M-1}}-M$, then the optimal $M^*$ is the greatest integer that is smaller than the solution of \eqref{eq:Optimal_M_equality}.

\section*{Acknowledgement}
The authors would like to thank Dr. Jeffrey G. Andrews for his helpful comments.

\bibliographystyle{IEEEtran}
\bibliography{IEEEabrv,Ref}

\end{document}